\documentclass[12pt,preprint]{aastex}

\def\gtorder{\mathrel{\raise.3ex\hbox{$>$}\mkern-14mu
             \lower0.6ex\hbox{$\sim$}}}
\def\ltorder{\mathrel{\raise.3ex\hbox{$<$}\mkern-14mu
             \lower0.6ex\hbox{$\sim$}}}

\def\Msun{\>{\rm M_{\odot}}}
\def\Mjup{M$_{Jup}$}

\shorttitle{Limits to Substellar Objects Around White Dwarfs}
\shortauthors{Debes et al.}

\begin{document}
\title{Cool Customers in the Stellar Graveyard III:
 Limits to Substellar Objects around nearby White Dwarfs using CFHT}
\author{John H. Debes\altaffilmark{1,2},
Jian Ge\altaffilmark{3},Christ Ftaclas\altaffilmark{4}}

\altaffiltext{1}{Department of Astronomy \& Astrophysics, Pennsylvania State
University, University Park, PA 16802}
\altaffiltext{3}{Department of Astronomy, University of Florida, Gainesville, Fl 32611}
\altaffiltext{4}{Institute for Astronomy, University of Hawaii, Honolulu, HI 96822}
\altaffiltext{2}{Based on observations obtained at the Canada-France-Hawaii Telescope (CFHT) which is operated by the National Research Council of Canada, the Institut National des Science de l'Univers of the Centre National de la Recherche Scientifique of France, 
and the University of Hawaii.}

\begin{abstract}
Results from a groundbased high contrast imaging 
survey of thirteen nearby white
dwarfs for substellar objects is presented. We place strict upper limits
on the type of substellar objects present, ruling out the presence of 
anything larger than $\sim$14 M$_{Jup}$ for eight of the white dwarfs
at separations $>$19 AU and corresponding to primordial separations of 
$\sim$3-6~AU assuming adiabatic mass loss without tidal interactions.
  With these results we place the first
upper limit on the number of intermediate mass stars with brown dwarfs at 
separations $>$ 13 AU.  We combine these results with previous work to place 
upper limits on the number of massive Jovian ($>$ 10 M$_{Jup}$) planets in
orbit around white dwarfs whose progenitors spanned a mass range of 1-7 M$_{\odot}$.
\end{abstract}   


\section{Introduction}
White dwarfs (WDs), the end state of stellar evolution for $\sim$1-8 $\Msun$,
are a population of stars that potentially hold an important key to directly 
imaging extrasolar planets \citep{burleigh02}.  They present several advantages compared to 
main sequence stars for strategies that rely on high contrast imaging.  
Due to their dense nature, WDs have small radii and 
cooling atmospheres that translate to surface fluxes orders of magnitude
dimmer than their main sequence progenitors.  Since they are hotter than any 
putative substellar companion, the companion's flux peaks well into the 
Rayleigh-Jeans tail of the WDs emission.  These two factors allow a modest 
contrast difference between the white dwarf and any possible substellar 
companions.

This is the third paper in a series that take slightly different approaches
in the search to directly image a nearby extrasolar planet and place limits 
on what type of companions could be present around each WD.  The first, \citet[][hereafter DSW05a]{
debes05a}, looked specifically at the WD G 29-38 
but utilized 2MASS photometry, pulsational timing, Gemini North Telescope high
contrast imaging, and Hubble Space Telescope (HST) coronagraphic images
 to constrain the presence of planets and brown dwarfs 
at distances from 0.1 AU to 50 AU.  The second paper, \citet[][hereafter DSW05b]{debes05b}, relied on 2MASS data and 
HST data to once again constrain the presence of planets and brown dwarfs 
around DAZ WDs, WDs with hydrogen atmospheres and the presence of weak metal
line absorption.  We extend that search now to nearby white dwarfs with well
modeled ages and measured parallaxes from the ground.  For a majority of our
targets we used the sample of \citet{bergeron01}.

We search for planets and brown dwarfs
with imaging using the PUEO/KIR instruments on the Canada
France Hawaii Telescope (CFHT).  This initial survey is intended to demonstrate
that useful detections of substellar and planetary objects are possible with
a large enough sample of nearby white dwarfs from the ground. 

 In Section \ref{cfhtobs2} we present the
 observations we performed and in Section \ref{cfhtobs3} we discuss how
our observations were analyzed.  
In Section \ref{cfhtobs4} we present the candidate
companions we have discovered as well as any background objects that may 
be present.  Several of these companions can be ruled out
through the use of second epoch observations.  For one
WD, WD 0208+396, we provide second epoch information for a candidate discovered
in DSW05b.  In Section \ref{cfhtobs4} we determine the
limits to companions that we could have detected.  Finally in Section \ref{cfhtobs5}
we summarize our results and synthesize them with the results of previous work.

\section{Observations}
\label{cfhtobs2}
Observations of all the white dwarfs were taken during three trips to 
CFHT--the first on October 11-14 2003, the second
on April 1, 2004, and the final on September 29, 2004.
  Observations were taken primarily in the J band on the 
first run and in the H band on the second and third
run using the KIR instrument in
conjunction with PUEO, the wavefront curvature AO system \citep{rigaut98}.  Table \ref{tab:targs} shows the list of targets including their V magnitudes, 
masses, primordial masses, estimated ages,
 effective temperatures, and distances.  The mass, T$_{eff}$, and
cooling age came from either \citet{bergeron01} or \citet{liebert04}, with the
exception of WD 0501+527.  WD 0501+527's parameters come from \citet{finley97}.  The 
primordial mass (M$_{i}$) was calculated by a theoretical initial to final
mass function given by M$_i=10.4\ln[(M_{WD}/ \Msun)/0.49] \Msun$ \citep{wood92}.  The 
main sequence lifetime of the star was determined by $t_{MS}=10$M$_{i}^{-2.5}$
 Gyr \citep{wood92}.  Since these relations only work well for M$_{WD}$ $> 0.54\Msun$, WD 0501+527's total age is unknown.

The very advantage these targets have for detecting planets
is offset by the fact
that most current AO systems cannot reliably correct atmospheric turbulence
for such faint objects.  With most large telescope AO systems requiring targets
with V $\ltorder$ 13, most of these targets would have to be imaged without the
help of AO.  However, the curvature wavefront sensor AO system of PUEO
 provides a 
heightened advantage by being able to guide on targets with V$\ltorder$16, 
allowing most nearby WDs to be accessible to AO correction \citep{rigaut98}.
  AO correction is
particularly useful for gaining spatial resolution as well as sensitivity 
against the near-IR background, the wavelength at which cool substellar 
objects become observable with current telescopes.
  These two benefits allow the more modestly
sized CFHT to compete realistically with larger telescopes in this area
without AO, as
well as with space based near-IR imaging.

Table \ref{tab:obs} shows our list of observations as well as total
integrations for each target WD.  
Most objects were observed for $\sim$1 hr using 240s sub-exposures that were
dithered in a 5\arcsec\ five point grid pattern for background
subtraction.  This left a $\sim$20\arcsec$\times$20\arcsec\ 
field of high sensitivity.  WD 1213+568 and WD
1633+572  had shorter total exposure times, with 15 minutes and
16  minutes respectively.  WD 0208+396, WD 0501+527, and WD 2341+321
had longer integrations of 90 minutes, 66 minutes, and 78 minutes, respectively.  During these observations high levels of relative humidity forced the telescope to be shut down for short periods of time, which resulted in a repetition
of some positions
of our dither pattern on the sky. 
  Objects that threatened to saturate the detector
had shorter sub-exposures.  This was the case for WD 1213+568 and WD 2140+207
whose sub-exposures were 60s and 120s respectively.
Flatfields were taken at the
beginning of each night.

As can be seen from Table \ref{tab:targs}, our targets ranged in brightness, which in turn
affected the performance of the AO correction.  Correction deteriorated towards
dawn on our second run 
as the sky background increased, and weather conditions varied throughout our
first run.  The third run had spectacular seeing throughout most of the
night (0.5\arcsec-0.6\arcsec\ in V), allowing diffraction limited images to be taken
 of WD 2140+207,
WD 2246+223, and WD 2341+321.  Throughout much of the second run, when most of our targets were
taken, the 
full width half maximum (FWHM) of our final images ranged from $\sim$140 
milliarcseconds to $\sim$200 milliarcseconds, compared to a diffraction limited
FWHM of 120 milliarcseconds.  WD 0501+527's final FWHM was 132 milliarcseconds,
 compared to the J band diffraction limited FWHM of 90 milliarcseconds.

\section{Data Analysis}
\label{cfhtobs3}
All data were flatfielded, background subtracted, registered, and combined into
final images.  These final images were used for two purposes: for deep
background limited imaging far from the central target star and as point 
spread function (PSF) 
reference stars for other observations.  Due to dithering, the highest 
sensitivity was generally within 7\arcsec\ of the target star. 

\begin{figure}[btp]
\begin{center}
\scalebox{0.60}{
\includegraphics{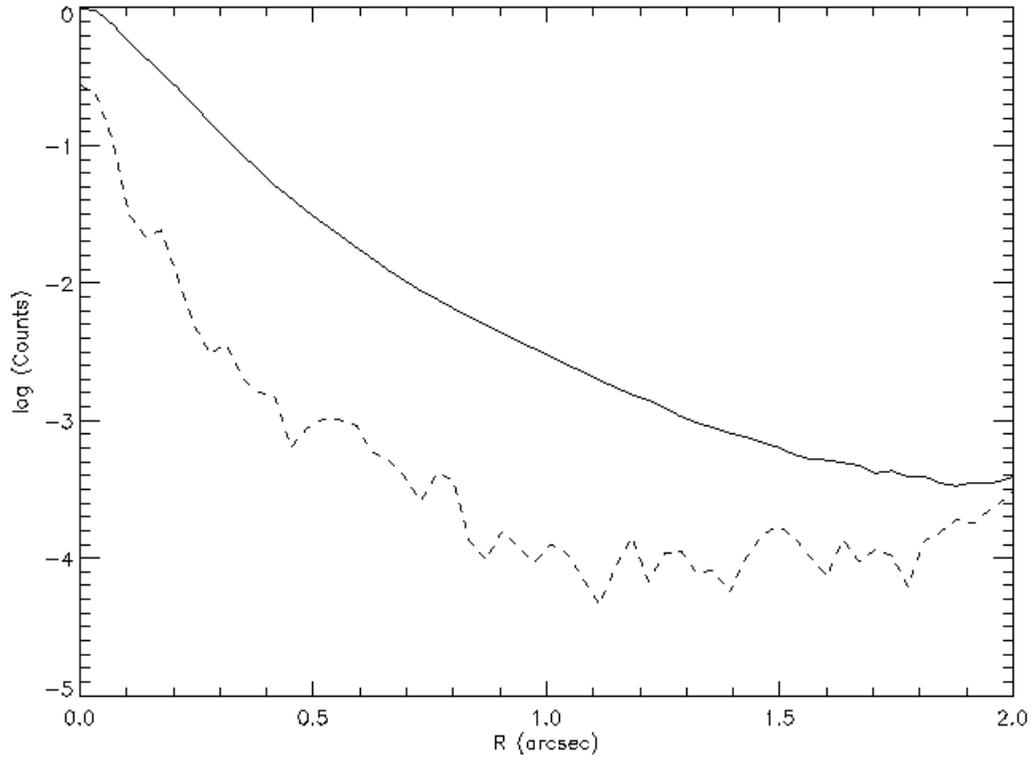}}
\caption{\label{fig:compcfht} Two azimuthally averaged PSFs for WD~1953-011, before subtraction 
(solid line) and after subtraction (dashed line).  This WD
had a contemporaneous PSF reference in the field which was used for subtraction
purposes.}
\end{center}
\end{figure}

In order to gain contrast close to each target white dwarf, we also employed
PSF subtraction to get high contrast to within 1\arcsec.  To achieve good 
results,
each registered sub-exposure was subtracted from another reference PSF image;
preferably from a reference that was brighter than the target and that had
a similar FWHM.  The subtraction
 images were median combined to produce the
final subtracted image.
In the case 
of WD 1121+216 and WD 1953-011,  
there was a brighter star in the field and that was used as a 
simultaneous reference.  Even though observations were separated by timescales
on the order of hours, we were able to get subtraction that
suppressed the PSF by 3-4 magnitudes
 at 0.8\arcsec (see Figure \ref{fig:compcfht}),
 with a higher sensitivity typically achieved in
the non-subtracted images beyond 2\arcsec. PSF subtraction was not possible for
WD 1213+528, WD 0208+396, and 
WD0521+527, since no suitable reference was available.
Figure \ref{fig:compcfht} shows a comparison before and after
PSF subtraction with a contemporaneous reference for WD 1953-011.

Any point sources that were detected had their flux measured by adding 
the counts within an aperture comparable to the FWHM of the particular image and
comparing the counts in the same size aperture with the target star.  A 
differential H magnitude was computed and then added to the 2MASS H magnitude
of the WD, taking into account the transformation from the 2MASS system to the
MKO system. Since AO PSFs tend to 
vary with time photometric accuracy is limited by this variation and we 
found it preferable to use differential magnitudes since to zeroth order
all PSFs in an image should be varying in the same manner.  The 
large isoplanatic patch of PUEO makes this a reasonable assumption \citep{rigaut98}.   Extended objects were interpreted to be 
background galaxies and had their total flux measured within a 
0.5\arcsec\ radius aperture and compared to the flux of the target star in
the field.  Typically, most of the light from a star was captured within a 
1.5\arcsec\ radius aperture, such that larger apertures changed the 
instrumental magnitude by $\sim$0.01 mag or less.  

\section{Candidate Companions and Background Objects}
\label{cfhtobs4}
Many targets showed nothing besides the primary in the field of view.  However,
six of the targets had other objects in the field which we designated as
potential candidates. Any candidate would have to be unresolved.  Where second
 epoch images were available, we used them to
determine if any candidate was co-moving with the primary.  If any second 
epoch images showed no common proper motion that candidate was eliminated.
Two candidates do not have second epoch information and remain as viable 
brown dwarf candidates.  Several of the higher latitude targets also had
nearby resolved galaxies within 10\arcsec, which we note 
in case they are useful for future groundbased study; such as with laser
guided AO or multi-conjugate AO.  
Table \ref{tab:gals} gives their positions and H band 
magnitudes within a 0.5\arcsec\ aperture.  One object, DSW 1, has already been
presented in DSW05b, but here we add its MKO H magnitude from our CFHT 
observations.

\subsection{WD 2341+321}
\begin{figure}[btp]
\begin{center}
\scalebox{0.40}{
\includegraphics{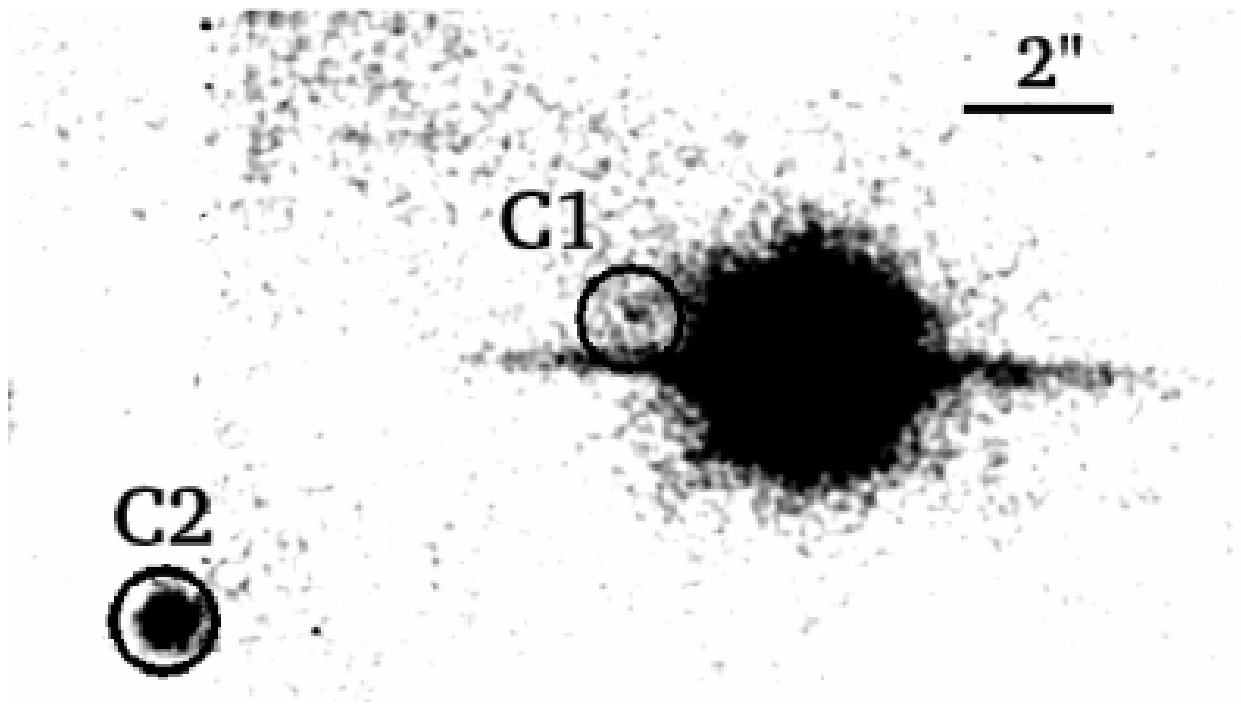} \hfill
\includegraphics{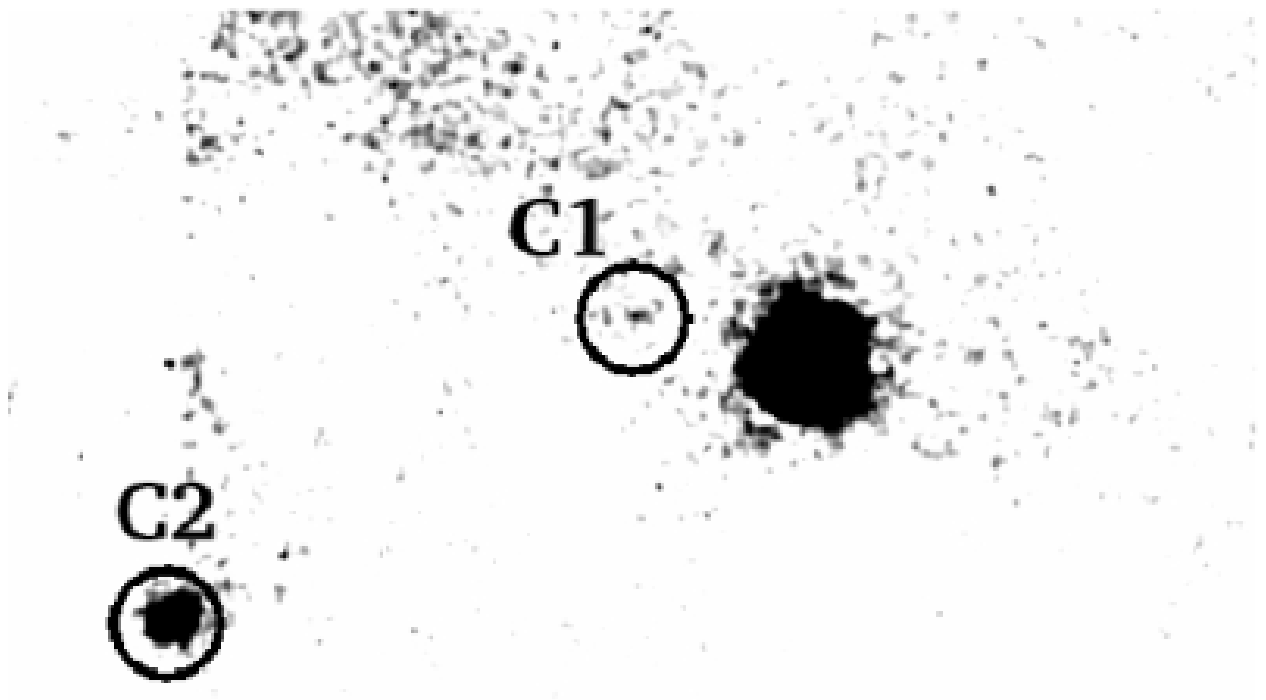}}
\caption{\label{fig:wd23} Candidate companion (C1) at a separation of 2.25\arcsec.
If this object is physically associated it would be an 11$M_{Jup}$ object.}
\end{center}
\end{figure}

WD~2341+321 has two candidate point sources--C1 and C2--that
 cannot be refuted with second
epoch POSS images.  Both are too faint to have been detected.  C1
is at an R of 9.17\arcsec$\pm$0.01 and a PA of 116$^\circ \pm$1, with an H magnitude of 18.5.  C2 is
detected closer in, after PSF subtraction.  Figure \ref{fig:wd23} shows the
original image and after PSF subtraction.  This dimmer candidate is more 
promising since it is closer to the target WD,
 and is detected at a S/N of 7 with an H magnitude of 22.3.  It has an R of 2.25\arcsec\ and PA of 72.5$^\circ \pm$1.  If both
are physically associated with WD 2341+321, they would be 27 \Mjup\ and
 13 \Mjup\
respectively.  At a distance of 16.6~pc, they would have orbital separations
of 37~AU and 152~AU corresponding to primordial separations of 13~AU and 54~AU,
given a current WD mass of 0.57$\Msun$ and an inferred initial mass
of 1.6$\Msun$.  However, they cannot be ruled associated until they demonstrate
common proper motion with WD 2341+321.  WD 2341+321's 
proper motion is 0.21\arcsec/yr,
 so it should
be relatively easy to determine common proper motion within a year \citep{hip}.
 
\subsection{WD 1121+216}

\begin{figure}[tbp]
\begin{center}
\scalebox{0.70}{
\includegraphics{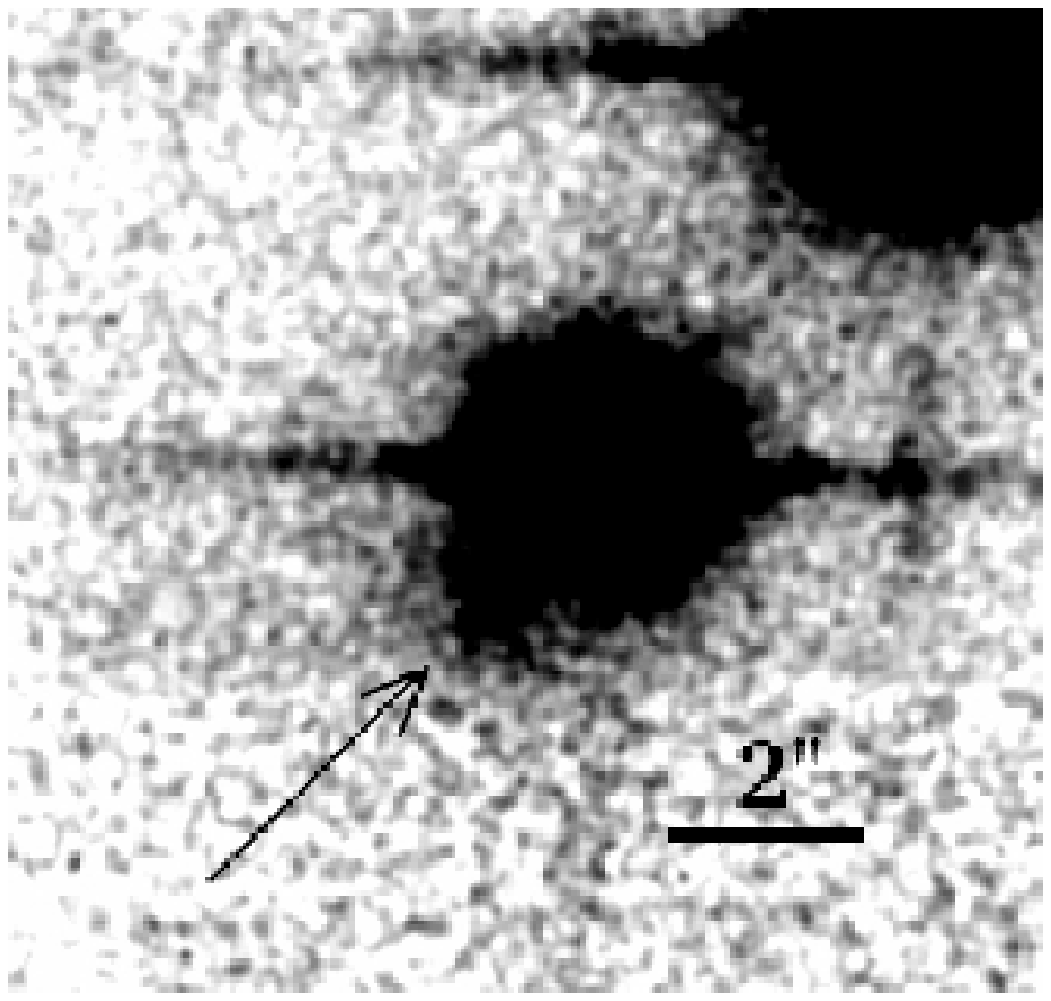}\hfill
\includegraphics{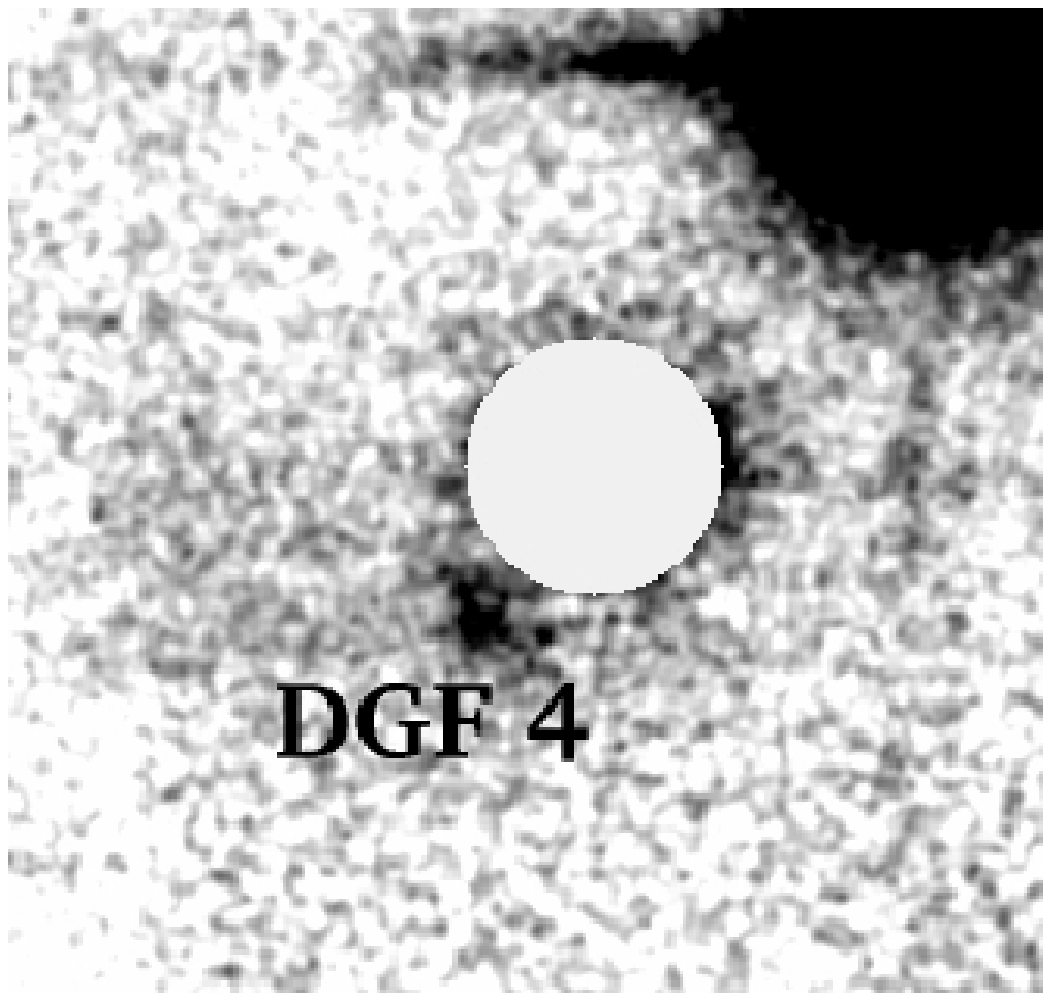}}
\caption{\label{fig:wd11} These images show extended emission discovered around
WD 1121+216 before (left panel) and after (right panel) PSF subtraction.  
Second epoch POSS images show
that it is a background galaxy.  The scale bar in the left panel represents 2\arcsec.}
\end{center}
\end{figure}

WD 1121+216 has a brighter star $\sim$5\arcsec\ away.  Inspection
of POSS plates clearly shows that it is a relatively fixed background
star and it is not a common proper motion 
companion.  
After PSF subtraction, WD 1121+216 shows emission that at first glance
 appears to be a dust disk or blob $\sim$ 20 AU from the WD.  Figure
\ref{fig:wd11} shows the emission.  It is clearly visible both in the 
original image and after PSF subtraction.  Inspection of the POSS 2 B plate
 shows that it is most likely 
a background galaxy, as there is an extended source
at the position of the emission currently seen near the WD.  Caution should be
taken with high latitude objects that appear to show extended emission as
a background galaxy can be mistaken for circumstellar emission.  Any such 
discovery should show common proper motion to be credible.  The background 
galaxy has a surface brightness of 20.1 mag/$\square$\arcsec.
  This detection demonstrates that
we could have discovered any circumstellar emission for our targets at 
approximately this level. 

\subsection{WD 1213+528}
WD 1213+528 shows a candidate companion $\sim$8\arcsec\ to the south, but 
inspection of POSS 2 plates shows that this object is not a common proper 
motion companion.

\subsection{WD 1953-011}
\begin{figure}[btp]
\begin{center}
\scalebox{0.70}{
\includegraphics{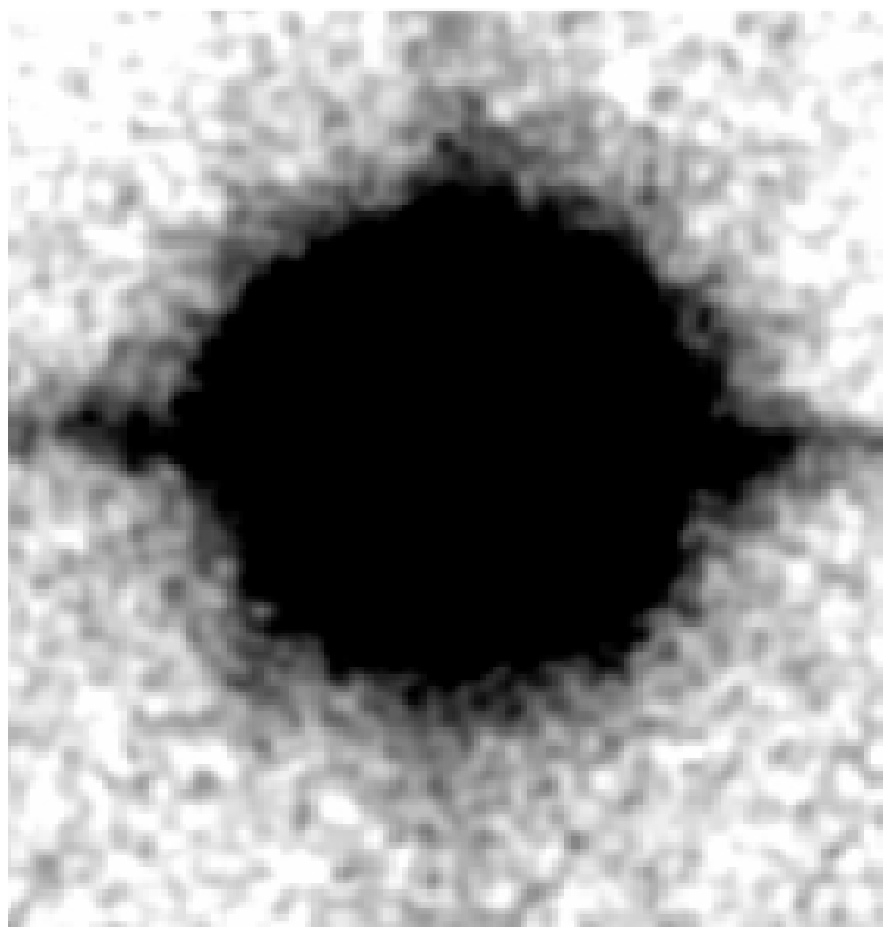} \hfill
\includegraphics{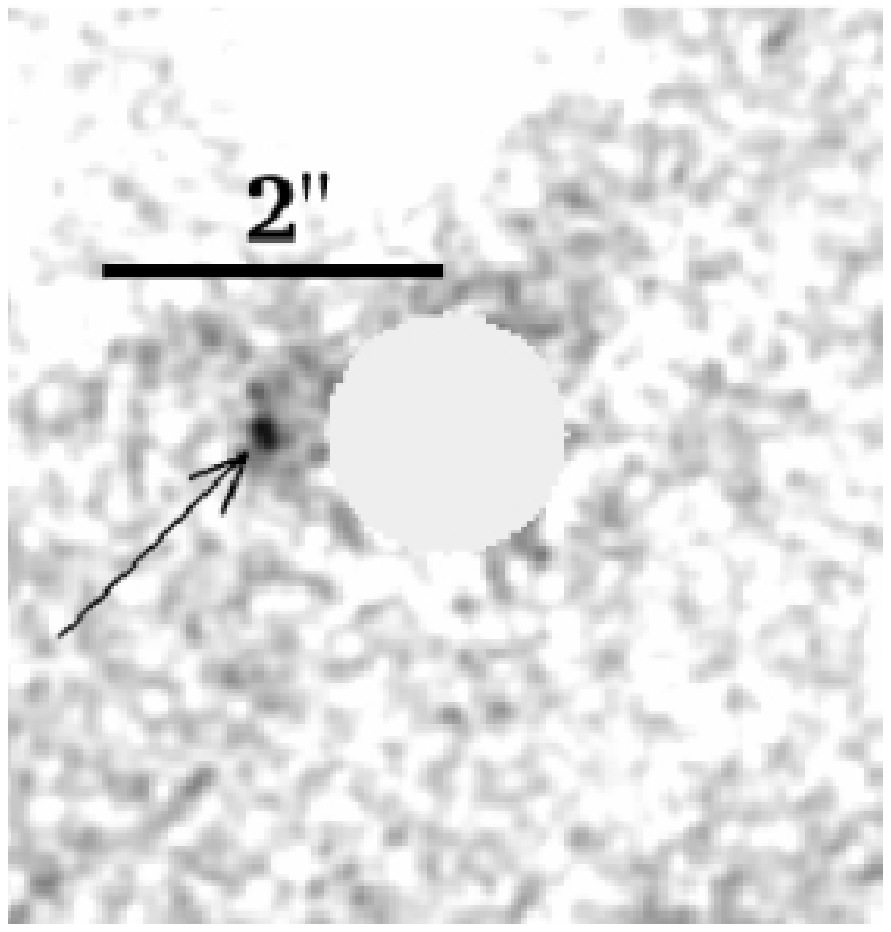}}
\caption{\label{fig:wd19} Images of a nearby background star near WD 1953-011
that show a
companion with $\Delta H$=7.6 with a separation of 1.08\arcsec\ before (left
panel) and
after (right panel) PSF subtraction.  This detection demonstrates the study's sensitivity to
point sources close to our targets.}
\end{center}
\end{figure}

WD 1953-011 has several nearby background sources, which are well 
separated.  Most are visible on POSS plates and due to WD 1953-011's proper
motion are easily discarded as possible proper motion companions.
The brightest background object in the field, $\sim$7\arcsec\ to the South, has
 a noticeable companion at a separation 1.08\arcsec $\pm$0.01, PA= 88.6$^\circ \pm$0.3
with a $\Delta H$=7.6.  
Figure \ref{fig:wd19} shows the star before and after PSF subtraction
and Gaussian smoothing.

A spectrophotometric SED using POSS B, V, R, and I magnitudes
and 2MASS J,H, and Ks magnitudes 
of the background star makes it consistent with 
either a $\sim$M0 dwarf at $\sim$300 pc or a K2 giant at $\sim$10 kpc
\citep{allen}.  If it is a main sequence star,
the companion would be a 0.07-0.08 $\Msun$ object according to the models
of \citet{baraffe03}.  If it is instead a giant,
the companion is an M dwarf.  The former explanation of 
a nearby M dwarf host star with a low mass companion
seems more plausible given
the low galactic latitude of the source and the apparent lack of 
significant reddening.  It is also possible
the two stars are not physically associated.  Despite the fact that this is
not relevant to our current study, this discovery demonstrates the efficacy of
our PSF subtraction technique.

\subsection{WD 2140+207}
WD 2140+207 has a dim, point-like object $\sim$5\arcsec\ away, with several 
point sources and galaxies in the surrounding field.  Most of the point sources can be
discriminated as background objects from POSS plates, including the near 
object discovered.  With the help of POSS PSF subtraction, a marginal detection
of the companion was possible on the POSS 2 B plate.  At epoch 1990.57, the 
time of the observations taken by POSS, the separation of the object had an 
R of 8.58\arcsec with a PA of 239.6$^\circ$ east of north.  In our CFHT 
observations the object had an R of 5.88 and a PA of 296.7$^\circ$.  This
is clearly a background object.

\subsection{WD 0208+396}
\begin{figure}[btp]
\begin{center}
\scalebox{0.5}{
\includegraphics{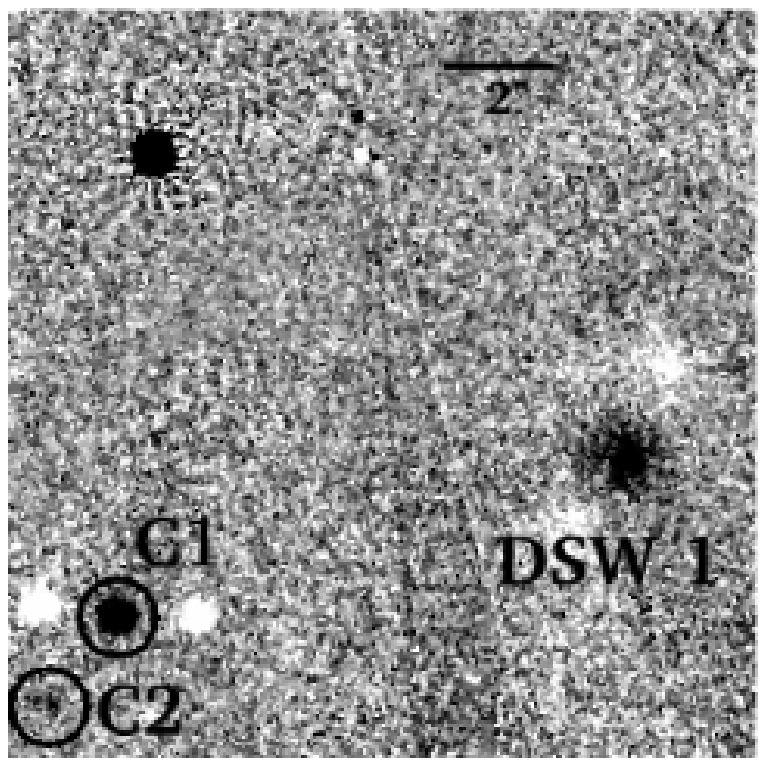} \hfill
\includegraphics{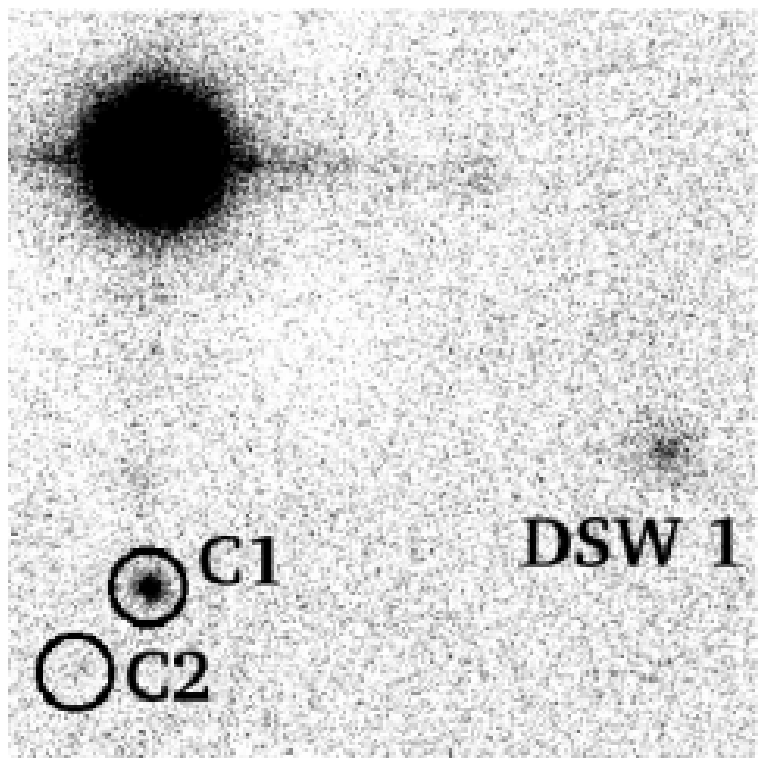}}
\caption{\label{fig:g74} A comparison between the HST and CFHT fields for 
WD 0208+395.  The HST field is on the left and the CFHT field is on the right.
The images are about 9\arcsec\ long on a side and shows one of the candidates
due South of WD 0208+396 which is masked.}
\end{center}
\end{figure}

Two candidate objects as well as a galaxy $\sim$8\arcsec\ away were discovered
in HST images presented in DSW05b.  The point source candidates were re-imaged
on our third visit to CFHT $\sim$1 year later and their H magnitudes measured.
  Figure \ref{fig:g74} shows the
images at the two epochs.  
C1 and C2 had H magnitudes of 19.35 and 22.22 respectively.
  In order to determine whether any of the 
candidates had common proper motion we needed to compare their positions 
relative to the HST observations in DSW05a.  In those observations the C1 was found to be at a separation
of 8.60$^\circ \pm$0.1 and a PA of 175$^\circ \pm$1.  Its F110W magnitude was
20.64$\pm$0.01.  WD 0208+396 has a proper motion of 1069 mas/yr in RA and -523 
mas/yr in Dec, which allows us to predict the position of the C1 if
it is not co-moving.  We predict C1's
 position with respect to WD 0208+396 to be $\Delta$RA=-0.41\arcsec$\pm$0.1
and $\Delta$Dec=-8.03\arcsec$\pm$0.1.  We find that the candidate is measured
at a position
 $\Delta$RA=0.03\arcsec\ and $\Delta$Dec=-8.02\arcsec.  C2
has an F110W magnitude of 23.5$\pm$0.1 and in the HST image had an
R=10.33\arcsec$ \pm$0.2 with a PA=169$^\circ \pm$2.  Its predicted position if
not co-moving was predicted to be $\Delta$RA=0.82\arcsec$\pm$0.1 and 
$\Delta$Dec=-9.60\arcsec.  The measured relative position was $\Delta$RA=1.27\arcsec\ and $\Delta$Dec=-9.51\arcsec.  There is a systematic, significant 
difference between the predicted $\Delta$RA and that measured for both 
candidates which are also spatially close.  Measurements of the relative 
position of the galaxy in the field also shows a similar 
discrepancy in where its relative position should be (it's obviously
not co-moving) which
supports the explanation that the CFHT
field is rotated clockwise by $\sim$1.7$^\circ$, which places all of the 
measured positions within the errors of the predicted positions.  Therefore,
we can state with certainty that the candidates are both background
objects.  

\section{Limits to Companions}
\label{cfhtobs5}
Since many targets did not have any possible companions, it is instructive to 
place limits on what kind of objects could be detected around each target.  We can place limits both for resolved and unresolved companions
by the combination of our imaging results and the measurement of these objects'
measured flux in comparison with their expected flux.

\subsection{Imaging}

\begin{figure}[tpb]
\begin{center}
\scalebox{0.5}{
\includegraphics{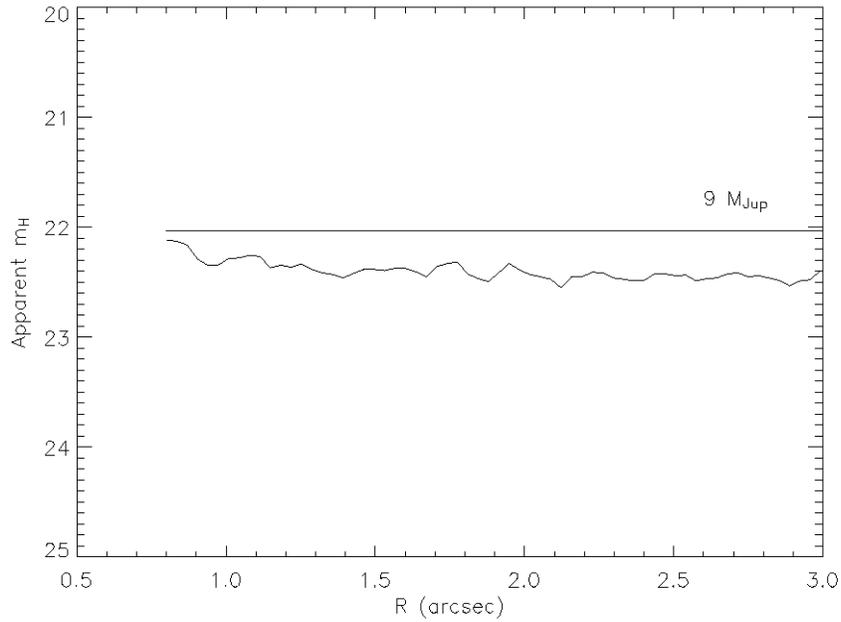}}
\caption{\label{fig:cfhtsens} Sample 5$\sigma$ sensitivity curve of WD 2246+223.
Overplotted is the H magnitude of a 9 M$_{Jup}$ companion at the age and 
distance of the target.}
.\end{center}
\end{figure}

For our images, we followed the same strategy for determining our imaging sensitivity as
in DSW05a and DSW05b.  This strategy is to implant artificial companions
into our images and try to recover them at a S/N of 5, in order to test the
sensitivity of our observations.  The main difference for AO imaging is that the 
PSF is not stable, so we use a  version of our target WD PSF normalized to 
1 DN.
 The implant would be scaled by a value, placed
within the field and an aperture approximately equal to the implant's core FWHM
was used to determine the S/N.  If the S/N was $>$5,
then the implant was considered recovered, otherwise the implant's scaled
value was increased.
  Values at 20 different angular locations
were determined at each radius, for azimuthal averaging.  The median of the 
different values was taken to give a final azimuthal averaged sensitivity.  
Relative photometry with respect to the target WD (or another unsaturated 
object in the field) was calculated and the 2MASS H magnitude for the target
WD was used to determine a final sensitivity.  Figure \ref{fig:cfhtsens} 
shows a typical sensitivity curve with PSF subtraction.

The values were then used with a grid of substellar spectral models to 
determine what kind of substellar object a limiting magnitude would correspond
to at the particular distance and age of the WD system.  Specifically, we used
the models of \citet{baraffe03}, primarily because they had isochrones that 
spanned the
mass range and age of interest to our target WDs.  The magnitudes were cross
checked with the models of \citet{bsl03} and 
for isochrones that overlapped they
provided similar results to within a magnitude or to within 1 or 2 M$_{Jup}$,
 thus giving us confidence that we could combine our 
results here with those in DSW05b.
  Using interpolation, we turned the 
observed sensitivities to specific masses at the particular ages of the WDs.
Table \ref{tab:sens} shows the final sensitivities for each WD.

\subsection{Near-IR Photometry}

While direct imaging is most sensitive to companions $>$1\arcsec\,
unresolved companions could still be present for some of these targets.  In
order to rule out companions at separations where imaging or PSF subtraction
could not resolve them, we turn to the near-infrared fluxes of these objects
provided by 2MASS photometry \citep{cutri03}.  Using the measured effective 
temperatures, gravities, and distances
 of the WDs given in the literature, we can model the expected
J, H, and Ks fluxes based on the models of \citet{bergeron95}.  If the photometry is of a high enough accuracy, one can place limits on the type of excesses
 present for
these objects.  These limits allow us to understand what types of companions
and dusty disks are ruled out.  
The details of this process have already been described
in DSW05a.
For our targets, the majority come from \citet{bergeron01}, but 
the WD 2341+321 parameters came from \citet{liebert04}.  DSW05b calculated the estimated 1 $\sigma$ limits for both samples in the J, H, and K$_{s}$ filters to be 0.04, 0.04, and 0.05 mag respectively 
for the \citet{bergeron01} sample. 
For the sample of  
\citet{liebert04} we found that the limits are 0.07, 0.10, and 0.15 for J, H, and K$_s$ respectively.  

The one exception is WD 0501+527, whose parameters are 
taken from \citet{finley97}.  The distance to WD 0501+527 is determined from
its Hipparcos parallax \citep{hip}.  In the \citet{finley97} sample, only 
spectroscopic properties were determined, so no attempt to model the distance 
was made.  Due to a lack of modeled distance, 
we cannot estimate the rough error in the modeling as an ensemble.
Rather, we compare $\Delta$J to the quoted photometric errors in 2MASS.  The
errors in J are $\sim$0.02, and since $\Delta$J falls within this range, we
use this as our estimate for a significant excess, which we determine to 
be an excess of 0.06 in J.  Since WD 0501+527 is so hot, its cooling time is
$\ll$1 Gyr and its total age depends entirely on its initial mass.  
Unfortunately this is unknown, so we calculate possible companion limits 
given a range of possible main sequence ages for this WD.

All of our other objects show no significant excess as well so 
 we need to determine to what mass limit we could have detected an
excess in our sample.  Taking the substellar models
 of \citet{baraffe03}, we took the 3$\sigma$ limits and interpolated between
the models to fit the estimated total ages of the white dwarf targets.
 We find that for all of our targets, any 
object more massive than $\sim$69 M$_{Jup}$ would have been detectable 
in the 2MASS search.  Therefore, all targets should not have any stellar 
companions present at close separations.  The exceptions to the limit are
WD 1213+568, which already has an unresolved companion M dwarf, and 
WD 0501+527, which is less sensitive due to its large T$_{eff}$.  Any further
excess beyond the companion of WD 1213+568 cannot be determined.  
Table \ref{tab:sens} shows
our results for unresolved and resolved companion sensitivities.  For the 
excess limits we take into account the distance to the WD to obtain a limit on the absolute magnitude of an object that could create an excess.

\section{Discussion}
\label{cfhtobs6}
We have surveyed thirteen white dwarfs for substellar objects.  From this 
search we have found two potential candidates both around the white dwarf
WD 2341+321.  This star requires follow-up observations to confirm or 
refute these candidates.  If any of the companions is confirmed to be 
co-moving, they
are dim enough to be consistent with substellar mass objects.  To date,
only two substellar objects are known to be in orbit around
nearby WDs \citep{zuckerman92,farihi04}.  With putative
 absolute magnitudes in the H band of $\sim$21-22, these would be hard to 
confuse with higher mass objects such as in young stellar populations 
\citep{mohanty04}.

To date, nine hydrogen white dwarfs with metal lines, so-called DAZs, have 
been searched for substellar objects--seven from our observations with HST
and two from the ground.
WD 1633+433 and WD 1213+529 both have been found to have small amounts of 
metals such as Ca in their atmospheres \citep{zuckerman03}.  WD 1213+529 has
an unresolved stellar companion, while WD 1633+433 appears from its 2MASS
photometry and our imaging to be devoid of anything $>$ 14 M$_{Jup}$ $>$ 15 AU
away and $>$ 48 M$_{Jup}$ at separations $<$ 15 AU.  Given that $\sim$25\% of DAs have
measurable metal lines and their explanation seems less likely due to 
ISM accretion and more due to unseen companions, either substellar or planetary,
they are interesting targets for faint companion searches \citep{zuckerman03,debes02}.  

Using a binomial type distribution that has been used to calculate
the frequency of brown dwarf companions to nearby stars, we can calculate limits to the frequency of substellar objects around DAZs as well as our full sample
of 20 white dwarfs \citep{mccarthy04}.  That distribution is given by
\begin{equation}
P(f,d)=f^d(1-f)^{N-d}\frac{N!}{(N-d)!d!}
\end{equation}
where $P$ is the probability, $f$ the true frequency of
objects, $N$ the number of observations, and $d$ the number
of successful detections.

If one integrates over all the probabilities, one can derive limits
that encompass 68\% of the distribution.  From these limits we can compare
 our results with both the radial velocity surveys and imaging surveys for 
brown dwarfs.  In this study we can place meaningful
limits to planet and brown dwarf formation around stars that originally
had masses between 1.5$\Msun$-7$\Msun$, the range of initial masses inferred
for our targets.  From our limit of $\sim$1\arcsec\ as the innermost separation
where we could have detected a companion for all of our targets, we can derive
an innermost projected orbital separation that we probed.  Since any companion
that is found today in an orbit with semi-major axis $a$ had a primordial
orbit M$_f$/M$_i$ times smaller before the
star lost its mass and turned into a white dwarf,
 we can probe inwards to regions that should have been sites for planet 
formation.  With a subset of our targets sensitive to planetary mass objects
at separations that would be where planets with orbits like Jupiter would be 
found we can study a region of parameter space complementary to radial velocity
surveys \citep{marcy00}.  Since our observations are also sensitive to
brown dwarfs we also complement surveys for widely separated brown dwarf
companions to white dwarfs \citep{farihi05}.

For our samples we neglect WD 0501+527 and WD 1213+529, since the observations
of these targets are 
significantly less sensitive than the other observations.  Of the 18 remaining
WDs from the DSW05b and CFHT samples, eight are DAZs and the rest are a mixture
of other white dwarf spectral types including DAs with no detectable metals in 
their atmospheres.  

In our DAZ sample, which includes those objects observed in DSW05b, WD 1633+433
and WD 1213+528,  the images of four WDs were sensitive enough to detect
planets and none were found.  Therefore, they do not
have planetary mass objects
$>$ 10 M$_{Jup}$ at projected separations $>$ 21 AU, corresponding to an
inferred minimum primordial separation of $>$ 6 AU.  When we integrate over
all possible probabilities we get an inferred limit of $<$20\%
for the frequency of massive planets in orbit around DAZs.  Assuming
that every DAZ may possess a planetary system, this is within a factor of 4
to the frequency of massive planets discovered with the radial velocity 
surveys 
with M$\sin(i)$ $>$ 10M$_{Jup}$, where 6 of 118 discovered planetary 
systems\footnote{http://www.obspm.fr/encycl/encycl.html} possess such
 companions $<$5-6 AU as well as radial velocity surveys of G giants \citep{marcy00,sato03}.  
Furthermore, none of the eight apparently single DAZs showed the presence of 
companions $>$70 M$_{Jup}$ in close, unresolved orbits.  This implies that
$<$12\% of DAZs have companions that are stellar.  Any unseen
object that could pollute a WD would have to be substellar for the majority
of current apparently single DAZs.

For our total sample of 18 WDs, we can also place limits on any object $>$
19 M$_{Jup}$ present at projected separations $>$ 34 AU and corresponding to
a minimum primordial orbit of $>$ 10 AU.  From zero detections in this sample,
we infer that intermediate mass stars from between 1.5-7$\Msun$ have
brown dwarf companions $<$6\% of the time.  Also, for our entire sample of 
nearby white dwarfs, seven of the eighteen were sensitive enough to detect 
massive planets (M$>$ 10 M$_{Jup}$) at projected separations of $>$ 21~AU, with
inferred primordial separations of $>$6~AU.  Therefore, the upper limit for the
presence of massive planets around intermediate mass stars is $<$13\%.  If we 
also include the results of a planet search amongst single white dwarfs in
the Hyades, we can effectively double our sample size of targets sensitive 
to massive planets \citep{friedrich05}.  In the \citet{friedrich05} survey,
they found no planets $>$ 10 M$_{Jup}$ at separations $>$23~AU, corresponding 
to primordial separations of $\sim$5~AU.  Combined with our results, the upper
limit for massive planets around single white dwarfs
(at 68\% confidence) is then closer to 7\%.

High spatial resolution imaging of white dwarfs will also
be important as supporting observations for Spitzer observations of white 
dwarfs that are looking for mid-IR excesses due to
substellar companions.  WD 1121-216 in particular may 
falsely show an excess due to it temporarily being coincident with a background
galaxy.  Approximately 100 WDs have been approved to be observed with Spitzer
in the Cycle 1 GO programs.  A combination of the Spitzer photometry and 
imaging would provide a more sensitive test for unresolved companions while
providing a check against source confusion due to Spitzer's larger PSF with
the IRAC camera, for example
\citep{fazio04}.  A large survey like that would also start placing rigorous
 limits
on the presence of faint companions to nearby white dwarfs.

\acknowledgements
JD and JG acknowledge partial support by NASA with grants NAG5-12115, NAG5-11427, NSF with grants AST-0138235, and AST-0243090. CF acknowledges support from
 NSF with grant AST-9987356.

\bibliography{g29bib}
\bibliographystyle{apj}

\begin{deluxetable}{lccccccc}
\tablecolumns{8}
\tablewidth{0pc}
\tablecaption{\label{tab:targs} List of WD Targets}
\tablehead{
\colhead{WD} & \colhead{V} & \colhead{M$_f$} & \colhead{M$_i$}&  \colhead{T$_{eff}$} &  
\colhead{D} & \colhead{Total Age} & \colhead{References} \\
 & & \colhead{($\Msun$)} & \colhead{($\Msun$)} & \colhead{(K)} & \colhead{(pc)}
& \colhead{(Gyr)} }
\startdata
0208+396 & 14.5 & 0.60 & 2.1 & 7310 & 16.7 & 2.9 & 1 \\
0501+527 & 11.8 & 0.53 &  & 61000 & 68.8 & & 2 \\
0912+536 & 13.8 & 0.75 & 4.4 & 7160 & 10.3 & 2.8 & 1 \\
1055-072 & 14.3 & 0.85 & 5.7 & 7420 & 12.2 & 3.0 & 1 \\
1121+216 & 14.2 & 0.72 & 4.0 & 7490 & 13.4 & 2.2 & 1 \\
1213+528 & 13.3 & 0.64 & 2.8 & 13000 & 38.6 & 1.0 & 3,4 \\
1334+039 & 14.6 & 0.55 & 1.2 & 5030 & 8.2 & 10.2 & 1 \\
1626+368 & 13.8 & 0.60 & 2.1 & 8640 & 15.9 & 2.6 & 1 \\
1633+433 & 14.8 & 0.68 & 3.4 & 6650 & 15.1 & 2.8 & 1 \\
1633+572 & 15.0 & 0.63 & 2.6 & 6180 & 14.4 & 3.8 & 1 \\
1953-011 & 13.7 & 0.74 & 4.3 & 7920 & 11.4 & 1.9 & 1 \\
2140+207 & 13.2 & 0.62 & 2.4 & 8860 & 12.5 & 2.1 & 1 \\
2246+223 & 14.4 & 0.97 & 7.1 & 10330 & 19.0 & 1.7 & 1 \\
2341+321 & 12.9 & 0.57 & 1.6 & 12570 & 16.6 & 3.4 & 3\\
\enddata
\tablerefs{(1) \citet{bergeron01} (2) \citet{finley97} (3) \citet{zuckerman03}
(4) \citet{bleach00}}
\end{deluxetable}

\begin{deluxetable}{lccc}
\tablecolumns{4}
\tablewidth{0pc}
\tablecaption{\label{tab:obs} Observations}
\tablehead{ \colhead{WD} & \colhead{Date(UTC)} & \colhead{Filters} & \colhead{Total
Integration(s)}}
\startdata
0208+396 & 11$\colon$10$\colon$46 2004-09-30 & H & 5280 \\
0501+527 & 14$\colon$22$\colon$43 2003-10-11 & J & 3840 \\
 & 15$\colon$12$\colon$45 2003-10-12 & K & 1920  \\
0912+536 & 06$\colon$33$\colon$58 2004-04-02 & H & 3600 \\
1055-072 & 08$\colon$12$\colon$16 2004-04-02 & H & 3600 \\
1121+216 & 09$\colon$24$\colon$38 2004-04-02 & H & 3600 \\
1213+528 & 10$\colon$47$\colon$20 2004-04-02 & H & 900 \\
1334+039 & 11$\colon$32$\colon$37 2004-04-02 & H & 3600 \\
1626+368 & 12$\colon$46$\colon$04 2004-04-02 & H & 3600 \\
1633+433 & 14$\colon$13$\colon$36 2004-04-02 & H & 3600 \\
1633+572 & 15$\colon$22$\colon$16 2004-04-02 & H & 960 \\
1953-011 & 05$\colon$09$\colon$33 2004-09-30 & H & 3600 \\
2140+207 & 06$\colon$27$\colon$51 2004-09-30 & H & 3600 \\
2246+223 & 07$\colon$44$\colon$44 2004-09-30 & H & 3600 \\
2341+321 & 09$\colon$56$\colon$37 2004-09-30 & H & 4680 \\
\enddata
\end{deluxetable}

\begin{deluxetable}{llll}
\tablecolumns{4}
\tablewidth{0pc}
\tablecaption{\label{tab:gals} Extragalactic Objects}
\tablehead{ \colhead{Name} & \colhead{RA} & \colhead{Dec} & \colhead{H} } 
\startdata
DSW 1 & 02 11 20.67 & +39 55 19.2 & 20.5 \\
DGF 1 & 10 57 34.75 & -07 31 22.8 & 19.8 \\
DGF 2 & 10 57 34.63 & -07 31 13.8 & 20.2 \\
DGF 3 & 10 57 35.15 & -07 31 11.5 & 20.5 \\
DGF 4 & 11 24 12.82 & +21 21 23.9 & 19.7 \\
DGF 5 & 21 42 40.97 & +20 59 49.5 & 19.0 \\
DGF 6 & 21 42 41.71 & +20 59 46.9 & 20.2 \\
DGF 7 & 22 49 05.83 & +22 36 37.3 & 19.0 \\
\enddata
\end{deluxetable}

\begin{deluxetable}{lcccc}
\tablecolumns{5}
\tablewidth{0pc}
\tablecaption{\label{tab:sens} Sensitivities}
\tablehead{
\colhead{WD} & \colhead{Excess Limit} & \colhead{Mass} & \colhead{Sensitivity
$>$1\arcsec} & \colhead{Mass} \\
 & \colhead{(M$_J$)} & \colhead{(\Mjup)} & \colhead{(H)} & \colhead{(\Mjup)}}
\startdata
0501+527 (1 Gyr) & 15.6 & 75 & 19.8 & 25 \\
 (5 Gyr) & & 80 &   & 42 \\
 (10 Gyr) & & 80 &  & 63 \\
0912+536 & 14.8 & 46 & 21.3 & 12 \\
1055-072 & 15.0 & 46 & 20.9 & 14 \\
1121+216 & 14.6 & 44 & 21.3 & 11  \\
1213+528 & - & - & 18.0 & 29 \\
1334+039 & 14.8 & 65 & 21.9 & 18 \\
1626+368 & 14.2 & 52 & 21.1 & 14 \\
1633+433 & 14.7 & 48 & 20.9 & 14 \\
1633+572 & 14.8 & 54 & 20.5 & 19 \\
1953-011 & 14.5 & 43 & 20.8 & 10 \\
2140+207 & 14.1 & 49 & 21.5 & 10  \\
2246+223 & 14.4 & 41 & 22.3 & 9  \\
2341+321 & 13.0 & 69 & 22.4 & 13 \\
\enddata
\end{deluxetable}

\clearpage

\clearpage
\clearpage

\end{document}